\documentclass[aps,pre,twocolumn,groupedaddress,showpacs,floatfix,superscriptaddress]{revtex4-1}

\usepackage{graphicx}
\usepackage{amsmath}
\usepackage{amsfonts}
\usepackage{amssymb}
\usepackage{epsfig}
\usepackage{color}
\usepackage{bbm}
\usepackage{nicefrac}
\usepackage{times}

\usepackage{graphicx}
\usepackage{amsmath}
\usepackage{amssymb}

\usepackage{color}
\usepackage[T1]{fontenc}
\usepackage[utf8]{inputenc}

\usepackage{hyperref}

\hypersetup{
 colorlinks=true,
citecolor=black,
linkcolor=black,
urlcolor=black
 }

\newcommand{\pd}[3]{\frac{ \partial^{ #1 } #2 }{ \partial #3^{ #1 } }}

\usepackage[colorinlistoftodos,prependcaption,textsize=small]{todonotes}

\begin{document}

\title{Multimodal stationary states under Cauchy noise}

\author{Micha{\l} Cie\'sla}
\email{michal.ciesla@uj.edu.pl} \affiliation{Marian Smoluchowski
Institute of Physics, and Mark Kac Center for Complex Systems
Research, Jagiellonian University, ul. St. {\L}ojasiewicza 11,
30--348 Krak\'ow, Poland}

\author{Karol Capa{\l}a}
\email{karol@th.if.uj.edu.pl} \affiliation{Marian Smoluchowski
Institute of Physics, and Mark Kac Center for Complex Systems
Research, Jagiellonian University, ul. St. {\L}ojasiewicza 11,
30--348 Krak\'ow, Poland}

\author{Bart{\l}omiej Dybiec}
\email{bartek@th.if.uj.edu.pl} \affiliation{Marian Smoluchowski
Institute of Physics, and Mark Kac Center for Complex Systems
Research, Jagiellonian University, ul. St. {\L}ojasiewicza 11,
30--348 Krak\'ow, Poland}

\date{\today}

\begin{abstract}
A L\'evy noise is an efficient description of out-of-equilibrium systems. The presence of L\'evy flights results in a plenitude of noise-induced phenomena.
Among others, L\'evy flights can produce stationary states with more than one modal value in single-well potentials.
Here, we explore stationary states in special double-well potentials demonstrating that a sufficiently high potential barrier separating potential wells can produce bimodal stationary states in each potential well.
Furthermore, we explore how the decrease in the barrier height affects the multimodality of stationary states. 
Finally, we explore a role of the multimodality of stationary states on the noise induced escape over the static potential barrier.
\end{abstract}


\maketitle

%%%%%%%%%%%%%%%%%%%%%%%%%%%%%%%%%%%%%%%%%%%%%%%%%%
%                                                %
%    BEGINNING OF TEXT                           %
%                                                %
%%%%%%%%%%%%%%%%%%%%%%%%%%%%%%%%%%%%%%%%%%%%%%%%%%

%%%%%%%%%%%%%%%%%%%%%%%%%%%%%%%%%%%%%%%%%%%%%%%%%%%%%%%%%%%%%%%%%%%%%%%%%%%%%%%%%%%%%%%%%
%%%%%%%%%%%%%%%%%%%%%%%%%%%%%%%%%%%%%%%%%%%%%%%%%%%%%%%%%%%%%%%%%%%%%%%%%%%%%%%%%%%%%%%%%
%%
%% introduction
%%
\section{Introduction\label{sec:introduction}}

Noise is one of the fundamental concepts in the theory of stochastic systems \cite{horsthemke1984}.
It is used to effectively approximate complex interactions of an observed particle with its environment.
After a long lasting dominance of a Gaussian paradigm, it has been documented and accepted that many situations significantly departs from the Gaussian approximation. 
In the non-equilibrium regime, noise can be of the $\alpha$-stable type \cite{dubkov2008} resulting in unexpected properties of noise-driven systems.

In the recent two decades, significant progress in the development of the theory of systems driven by the L\'evy noise has been achieved \cite{metzler2000,barkai2001,anh2003,brockmann2002,chechkin2006,jespersen1999,yanovsky2000,schertzer2001}.
L\'evy flights have been studied in various situations \cite{klages2008}. 
Their applications cover, among others, economy and finance \cite{bouchaud1990}, superdiffusion of micellar systems \cite{bouchaud1991}, studies of turbulence \cite{shlesinger1986b}, description of photons in hot atomic vapours \cite{mercadier2009levyflights}, laser cooling \cite{cohen1990,barkai2014} and therapeutic aspects \cite{cabrera2004}.
These studies cover experimental \cite{solomon1993,barthelemy2008} and various  theoretical aspects  \cite{eliazar2003,sokolov2004b,garbaczewski2009,garbaczewski2010} of L\'evy flights.

In the context of current research, the problem of stationary states in systems driven by the L\'evy noises is especially important.
Stationary states in a single-well potential can be bimodal what is well known and documented \cite{chechkin2002,chechkin2003,chechkin2004,chechkin2006,chechkin2008introduction}.
Moreover, conditions for a steepness of single-well potentials that can bound L\'evy flights have been developed \cite{dybiec2010d}.
Furthermore, in \cite{capala2018multimodal} we have presented sample, fine tailored, single-well potentials resulting in stationary states with an arbitrary number of modal values.
In \cite{capala2018multimodal} and \cite{chechkin2002,chechkin2004} the number of modal values has been attributed to the number of maxima of the potential curvature.

The multimodality of stationary states can be also dynamically induced \cite{dybiec2007c,calisto2017forced}.
For example in single-well potentials perturbed by the Gaussian white noise and the Markovian  dichotomous noise the stationary state can be bimodal.
For such a system the maxima of stationary densities are placed in the vicinity of minima of the altered potential \cite{dybiec2007c} indicating differences between mechanisms producing multimodal stationary states by solely action of the L\'evy noise and the combined action of the Gaussian white noise and the Markovian dichotomous noise.

Within the current manuscript we further inspect the problem of multimodal stationary states in systems driven by the L\'evy noises, however, we focus on double-well potentials.
Therefore, we explore a situation when the number of modal values in the stationary state is larger than the number of minima of the potential.
We demonstrate that, if the potential minima are deep enough, they are sufficient to produce multiple maxima of a stationary state in any of potential wells.

The studied model is presented in the next section (Sec.~\ref{sec:model}). The main results are included in Secs.~\ref{sec:dpotential} -- \ref{sec:hpotential} and Sec.~\ref{sec:escape}, while Sec.~\ref{sec:spotential} plays an introductory role. 
The paper is closed with Summary and Conclusions (Sec.~\ref{sec:summary}) and supplemented with Appendices~\ref{sec:app-n4} -- \ref{sec:app-numerical} presenting relevant technicalities.

%%%%%%%%%%%%%%%%%%%%%%%%%%%%%%%%%%%%%%%%%%%%%%%%%%%%%%%%%%%%%%%%%%%%%%%%%%%%%%%%%%%%%%%%%
%%%%%%%%%%%%%%%%%%%%%%%%%%%%%%%%%%%%%%%%%%%%%%%%%%%%%%%%%%%%%%%%%%%%%%%%%%%%%%%%%%%%%%%%%
%%
%% model
%%
\section{Model and Results\label{sec:model}}

We start with the basic information regarding overdamped stochastic dynamics in single-well potentials (Sec.~\ref{sec:spotential}). 
Afterwards, in Sec.~\ref{sec:dpotential}, using this information, we study various types of double-well potentials which are able to produce stationary states characterized by a larger number of modal values than the number of minima of the potential.
Finally, in Sec.~\ref{sec:hpotential}, we investigate a model with a varying barrier height which allows to increase understanding of the role played by the potential barrier separating minima of double-well potentials. 

The overdamped motion in an external potential $V(x)$ is described by the following Langevin equation
\begin{equation}
 \frac{dx}{dt}=-V'(x)+\sigma \zeta_\alpha(t),
 \label{eq:langevin}
\end{equation}
where $\zeta(t)$ is the symmetric $\alpha$-stable L\'evy type noise, i.e. the formal time derivative of the symmetric $\alpha$-stable motion \cite{janicki1994b}.
The parameter $\sigma$, in Eq. (\ref{eq:langevin}), might be interpreted as a noise strength.
Eq.~(\ref{eq:langevin}) is supplemented with the initial condition $x(0)=x_0$.
Accordingly, the stochastic process $\{X(t), t\geqslant 0\}$ governed by Eq.~(\ref{eq:langevin}) has increments
\begin{eqnarray}
\label{eq:discretization}
 \Delta x & = & x(t+\Delta t)-x(t) \\
 & = & -V'(x(t))\Delta t +\Delta t^{1/\alpha} \sigma \zeta_t \nonumber.
\end{eqnarray}
In Eq.~(\ref{eq:discretization}) $\zeta_t$ represents independent, identically distributed random variables  \cite{chambers1976,weron1995,weron1996} following the symmetric $\alpha$-stable density \cite{janicki1994,janicki1996} with the unity scale parameter and the characteristic function $\phi(k)$
\begin{equation}
 \phi(k)=\exp\left[ - |k|^\alpha \right].
 \label{eq:fcharakt}
\end{equation}
Within simulations we use $\sigma=1$, nevertheless, due to various types of considered potentials, the comparison with analytical results requires reintroduction of the scale parameter $\sigma$, see Appendix~\ref{sec:app-n4} and \ref{sec:app-n}.
The non-unity scale parameter is reintroduced by the multiplication of $\zeta_t$ by $\sigma$.
Such a multiplied (rescaled) random variable has the characteristic function
\begin{equation}
 \phi(k)=\exp\left[ - \sigma^\alpha |k|^\alpha \right],
 \label{eq:fcharakt2}
\end{equation}
which is the typical form of the characteristic function of symmetric $\alpha$-stable densities.
The stability index $\alpha$ ($0<\alpha \leqslant 2$) describes an asymptotic  power-law decay of $\alpha$-stable densities, which for $\alpha<2$ is of a $ |x|^{-(\alpha+1)}$ type. The scale parameter $\sigma$ controls the distribution width.
For $\alpha<2$, the variance of an $\alpha$-stable density is infinite, thus the distribution width can be defined by the interquantile width or fractional moments.
For $\alpha=2$, the characteristic function (\ref{eq:fcharakt2}) reduces to the characteristic function of the normal (Gaussian) distribution with the probability density
\begin{equation}
    f_2(x)=\frac{1}{\sqrt{4\pi\sigma^2}} \exp\left[ - \frac{x^2}{4\sigma^2}  \right].
\end{equation}
The case of $\alpha=1$ corresponds to the Cauchy distribution
\begin{equation}
    f_1(x)=\frac{\sigma}{\pi(x^2+\sigma^2)},
\end{equation}
which is extensively used here.
For clarity and practical reasons, the scale parameter is extracted from the noise, see Eq.~(\ref{eq:langevin}).
From Appendices~\ref{sec:app-n4} and~\ref{sec:app-n} it can be deducted how the scale parameter can be reintroduced.

The evolution of the probability density generated by Eq.~(\ref{eq:langevin}) is described by the fractional Smoluchowski-Fokker-Planck equation \cite{samorodnitsky1994,podlubny1998,yanovsky2000}
\begin{equation}
 \pd{}{p(x,t)}{t} = -\pd{}{}{x} V'(x,t)p(x,t) + \sigma^{\alpha} \pd{\alpha}{p(x,t)}{|x|}.
 \end{equation}
The operator $\partial^\alpha/\partial |x|^\alpha$ is the fractional Riesz-Weil derivative \cite{podlubny1998,samko1993} which can be  defined via the Fourier transform $
 \mathcal{F}_k\left( \frac{\partial^\alpha f(x)}{\partial |x|^\alpha} \right)=-|k|^\alpha \mathcal{F}_k\left(f(x)\right).$

Results included in following subsections have been constructed numerically by methods of stochastic dynamics.
Eq.~(\ref{eq:langevin}) was integrated by the Euler-Maryuama method with the time step of integration $\Delta t=10^{-5}$ and  $10^6$ -- $10^8$ repetitions.  
More details can be found in Appendix~\ref{sec:app-numerical}.

%%%%%%%%%%%%%%%%%%%%%%%%%%%
%
% s potential
%
\subsection{Single-well potentials\label{sec:spotential}}

First, we study the motion of a particle driven by the L\'evy noise in single-well potentials of
\begin{equation}
 V(x)=n x^n
 \label{eq:spotential}
\end{equation}
type, where $n$ is even and greater than 0.
Stationary states for potentials given by Eq.~(\ref{eq:spotential}) exist for the sufficiently large steepness (exponent) of the potential $n$ \cite{dybiec2010d}, i.e.
\begin{equation}
    n > 2 -\alpha.
\end{equation}
The steepness of the potential, which is sufficient to produce stationary states, depends on the stability index $\alpha$ characterizing asymptotics of random pulses. 
Therefore, the existence of a stationary state does not depend solely on the potential, but it is also sensitive to the stability index $\alpha$ characterizing the noise.
Furthermore, if a stationary state exists, it is not of the Boltzmann-Gibbs type \cite{eliazar2003}.

In Appendices~\ref{sec:app-n4} and~\ref{sec:app-n}, starting from known formulas  \cite{chechkin2002,chechkin2003,chechkin2004,chechkin2006,chechkin2008introduction}, it is demonstrated  that for $n=4$ with $\alpha=1$ the stationary solution of Eq.~(\ref{eq:langevin}) is
\begin{equation}
 p_{\alpha=1}(x) = 
\frac{4}{\pi (4\sigma)^{\nicefrac{1}{3}}} \times \frac{1}{\left[ \frac{4x}{(4\sigma)^{\nicefrac{1}{3}}}\right]^4-\left[\frac{4x}{(4\sigma)^{\nicefrac{1}{3}}}\right]^2+1}.
 \label{eq:quartic-stationary}
\end{equation}
The stationary density~(\ref{eq:quartic-stationary}) depends on the scale parameter $\sigma$.
Please note, that for a finite $n$, the stationary density is sensitive to the scale parameter $\sigma$ characterizing strength of noise pulses, see \cite{dubkov2007} and Appendix~\ref{sec:app-n}.

The infinite rectangular potential well is recovered in the $n\to \infty$ limit of Eq.~(\ref{eq:spotential}), see \cite{kharcheva2016}.
For the infinite rectangular potential well the stationary state is \cite{denisov2008}
\begin{equation}
p_{\infty}(x)=\frac{\Gamma(\alpha) (2L)^{1-\alpha}  (L^2-x^2)^{\alpha/2-1}}{\Gamma^2(\alpha/2)}.
 \label{eq:rec-stationary}
\end{equation}
In contrast to the finite $n$, the stationary density (\ref{eq:rec-stationary}) does not depend on the scale parameter $\sigma$.

\begin{figure}[!h]%
\centering
\begin{tabular}{c}
\includegraphics[angle=0,width=0.95\columnwidth]{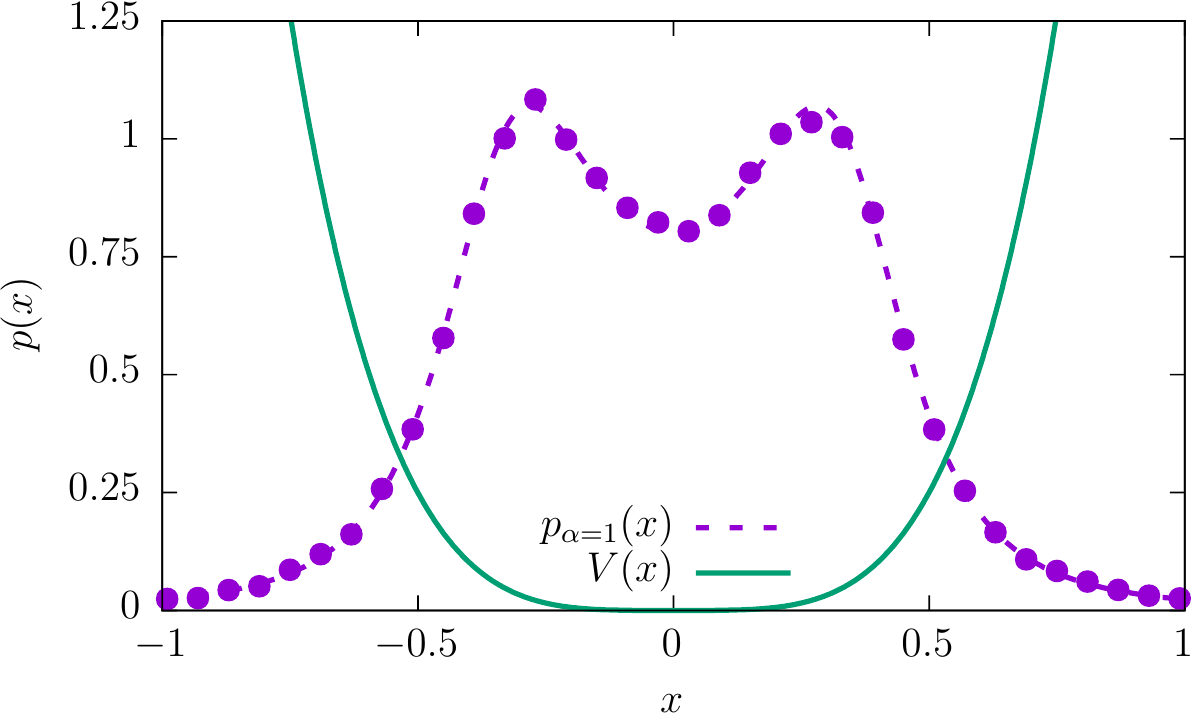} \\
\end{tabular}
\caption{The stationary state for a single-well potential with $n=4$ and $\alpha=1$.
Dashed line presents the theoretical density given by Eq.~(\ref{eq:quartic-stationary}), solid line represents the quartic $V(x)=4x^4$ potential. Points depict results of stochastic simulations.}
\label{fig:spotential}
\end{figure}

In order to verify the correctness of implemented numerical methods, see Appendix~\ref{sec:app-numerical}, we have performed a computer simulation for $V(x)=4x^4$. Fig.~\ref{fig:spotential} demonstrates the perfect agreement between results of a stochastic simulation (points) and the exact formula (dashed line) given by Eq.~(\ref{eq:quartic-stationary}).

Further numerical studies are devoted to the investigation of changes in the stationary densities in single-well potentials of $V(x)=nx^n$ type induced by the increasing exponent (steepness) $n$. 
Analytical results for such class of potentials can be obtained by a transformation of \cite[Eqs.~(38) and (39)]{dubkov2007}, see Appendix~\ref{sec:app-n}.
In Fig.~\ref{fig:spotential-n} these formulas are depicted by dashed lines and compared with results of stochastic simulations marked by  points.
Finally, the solid line presents the stationary density for the infinite rectangular potential well, see Eq.~(\ref{eq:rec-stationary}).
From Fig.~\ref{fig:spotential-n} it is clearly visible that results of stochastic simulations perfectly agree with analytical formulas.
Moreover, analogously like in \cite{kharcheva2016}, it is demonstrated that the stationary density approaches the stationary state for the infinite rectangular potential well in the limit of $n\to\infty$, see Eq.~(\ref{eq:rec-stationary}) and \cite{denisov2008}.
Due to symmetry of the potential (\ref{eq:spotential}) stationary states are also symmetric.
With the increasing $n$, modal values of the stationary density are shifted towards $x=\pm 1$.
Finally, in the limit of $n\to\infty$, modal values are located at impenetrable boundaries.
This test is crucial for the considerations performed in Sec.~\ref{sec:dpotential}, where double-well potentials based on Eq.~(\ref{eq:spotential}) are considered.

Multimodal stationary states are produced because of the interplay between the deterministic and random forces. The deterministic force $-V'(x)$ always acts towards the minimum of the potential $V(x)$.
The random force is the only factor which causes excursions towards large $|x|$.
Therefore, the emerging stationary density is the distribution that balances random excursions and deterministic sliding, which is interrupted by long jumps.
For the large exponent $n$, the potential $V(x)$ given by Eq.~(\ref{eq:spotential}) is very close to the infinite rectangular potential well. 
A random jump to the right or to the left moves the particle to the point where the restoring force is very large.
Consequently, the particle immediately slides to $|x|\approx 1$ and waits there for the next random pulse.
With the decreasing exponent $n$ the point where particle can slide down moves towards $x=0$.
At the same time, for $n\geqslant 2$, the time needed to reach the origin by a deterministic sliding is infinite.
The impossibility of reaching $x=0$ in a deterministic way in a finite time makes the stationary state bimodal.

\begin{figure}[!h]%
\centering
\begin{tabular}{c}
\includegraphics[angle=0,width=0.95\columnwidth]{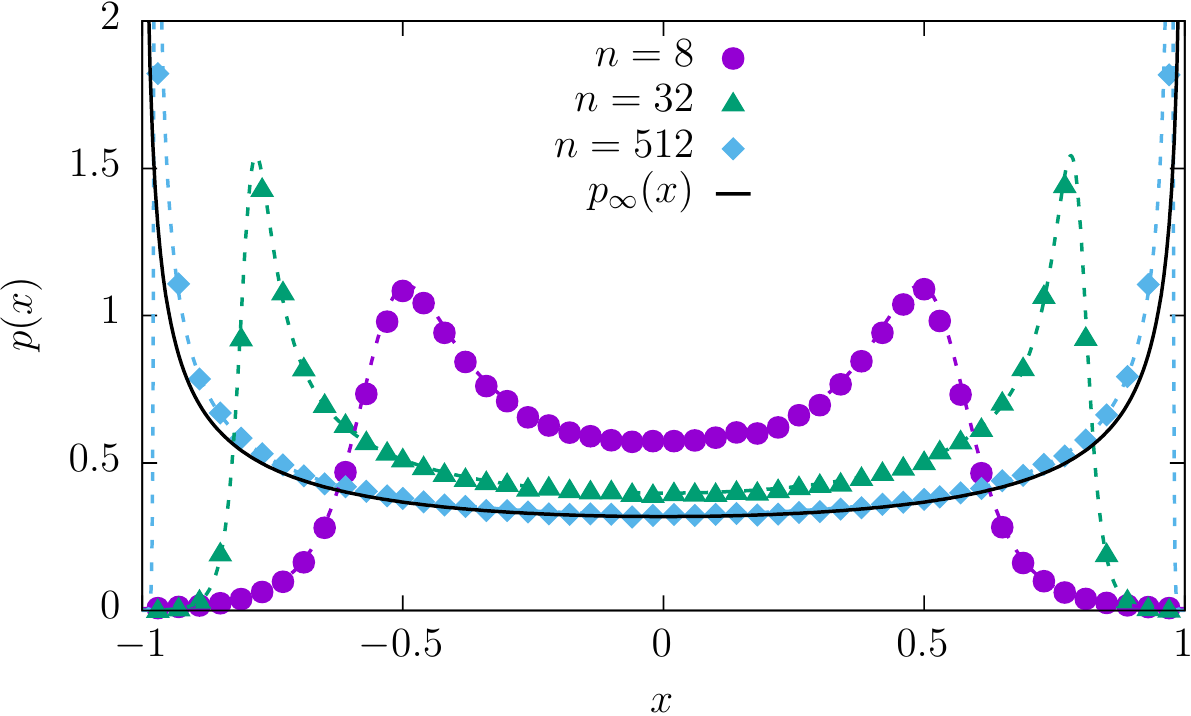} \\
\end{tabular}
\caption{Stationary states for single-well potentials $V(x)=nx^n$ for the increasing exponent $n$ with $\alpha=1$.
Dashed lines depict theoretical densities, see  \cite{dubkov2007}.
The solid line presents the theoretical $n\to\infty$ density given by Eq.~(\ref{eq:rec-stationary}).
}
\label{fig:spotential-n}
\end{figure}

%%%%%%%%%%%%%%%%%%%%%%%%%%%
%
% d potential
%
% \clearpage
\subsection{Double-well potentials\label{sec:dpotential}}

\begin{figure}[!h]%
\centering
\begin{tabular}{c}
\includegraphics[angle=0,width=0.95\columnwidth]{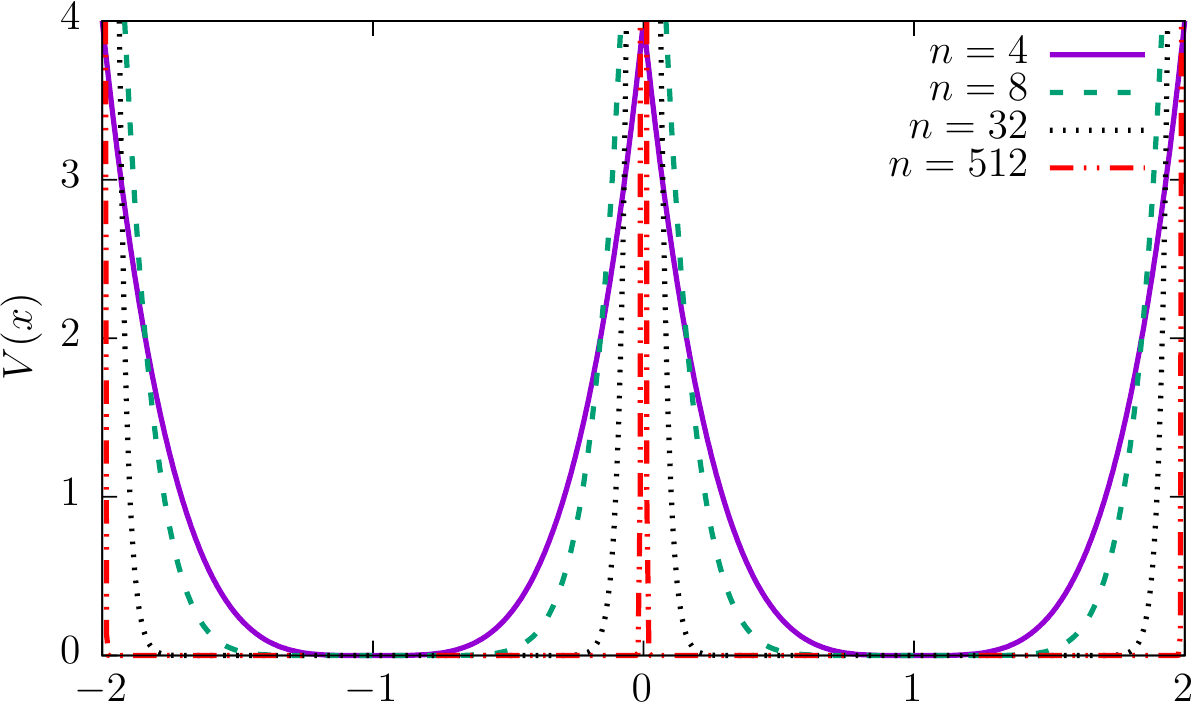} \\
\end{tabular}
\caption{Double-well potentials, see Eq.~(\ref{eq:dpotential}), with the increasing exponent $n$ ($n\in\{2,8,32,512$\}).
}
\label{fig:dpotential}
\end{figure}

As it is well known, and also demonstrated  in Sec.~\ref{sec:spotential}, it is possible to create a bimodal stationary state in a single well potential \cite{chechkin2002,chechkin2003,chechkin2004,chechkin2006,chechkin2008introduction}.
As the next step, it is intriguing to verify conditions necessary to produce stationary states with higher modality than the number of potential minima.
Moreover, it is interesting to inspect how the shape of the stationary state depends on the height of the potential barrier separating minima of a double-well potential.
In order to explore these issues, a special potential needs to be selected. 
Firstly, it has to behave like one of single-well potentials with $n \geqslant 2$ in the vicinity of its minima.
Moreover, potential wells must be wide enough to avoid interference of structures (e.g. peaks), emerging in different potential wells.
In addition, the potential barrier needs to be steep enough in order to produce internal minima of the stationary probability density.
Finally, in the limit of $n \rightarrow \infty$, the height of the potential barrier should be infinite.
The simplest choice fulfilling all required properties is a double-well potential based on the single-well potential given by Eq.~(\ref{eq:spotential}).
Therefore, we use the following double-well potential
\begin{equation}
 V(x)=n(|x|-1)^n=
 \left\{
 \begin{array}{ccl}
 n (x+1)^n & \mbox{for} & x<0 \\
 n (x-1)^n & \mbox{for} & x \geqslant 0 \\
 \end{array}
 \right.,
 \label{eq:dpotential}
\end{equation}
where $n$ is even and positive.
Sample double-well potentials given by Eq.~(\ref{eq:dpotential}) with various $n$ are depicted in Fig.~\ref{fig:dpotential}.
With the increasing exponent $n$ the potential $V(x)=n(|x|-1)^n$ approaches the sum of two infinite rectangular potential wells.
The height of the potential barrier is equal to $n$, i.e. $V(0)=n$ for every $n$.
Importantly, for a finite $n$, minima of the potential (\ref{eq:dpotential}) are located at $\pm 1$ regardless of the steepness exponent $n$.

\begin{figure}[!h]%
\centering
\begin{tabular}{c}
\includegraphics[angle=0,width=0.95\columnwidth]{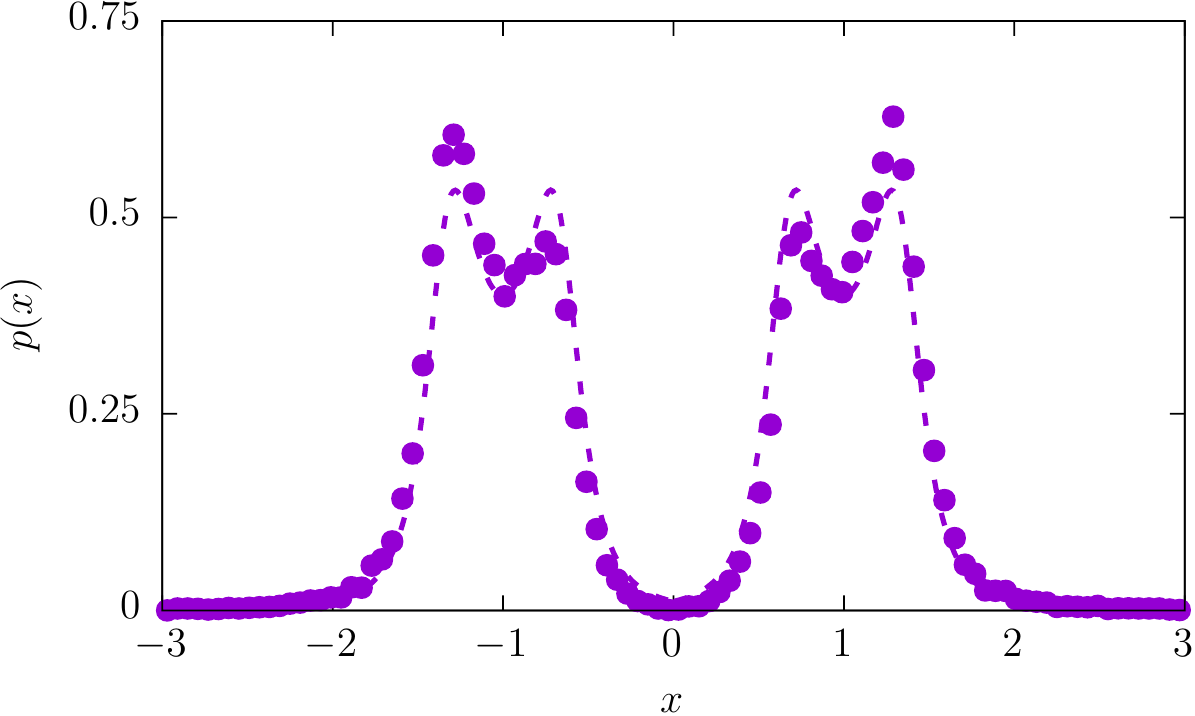} \\
\end{tabular}
\caption{The stationary density for  the double-well potential (\ref{eq:dpotential}) with $n=4$ and $\alpha=1$.
The dashed line depicts the probability density given by Eq.~(\ref{eq:sumofshifted}).
}
\label{fig:twowells4}
\end{figure}

Figure~\ref{fig:twowells4} shows results of a stochastic simulation for the double-well potential~(\ref{eq:dpotential}) with $n=4$.
The dashed line presents
\begin{equation}
 p(x)=
 \left\{
 \begin{array}{ccl}
  p_{\alpha=1}(x+1)/2 & \mbox{for} & x <0 \\
  p_{\alpha=1}(x-1)/2 & \mbox{for} & x >0 \\
 \end{array}
 \right., 
 \label{eq:sumofshifted}
\end{equation}
where $p_{\alpha=1}(x)$ is given by Eq.~(\ref{eq:quartic-stationary}).
The exact density (points) differs from $p(x)$ given by Eq.~(\ref{eq:sumofshifted}), nevertheless these differences are not so pronounced.
These differences originate in the finite height of the barrier separating the left and the right minimum of the double-well potential.
Therefore, internal maxima of the stationary density are lower than the outer one.

\begin{figure}[!h]%
\centering
\begin{tabular}{c}
\includegraphics[angle=0,width=0.95\columnwidth]{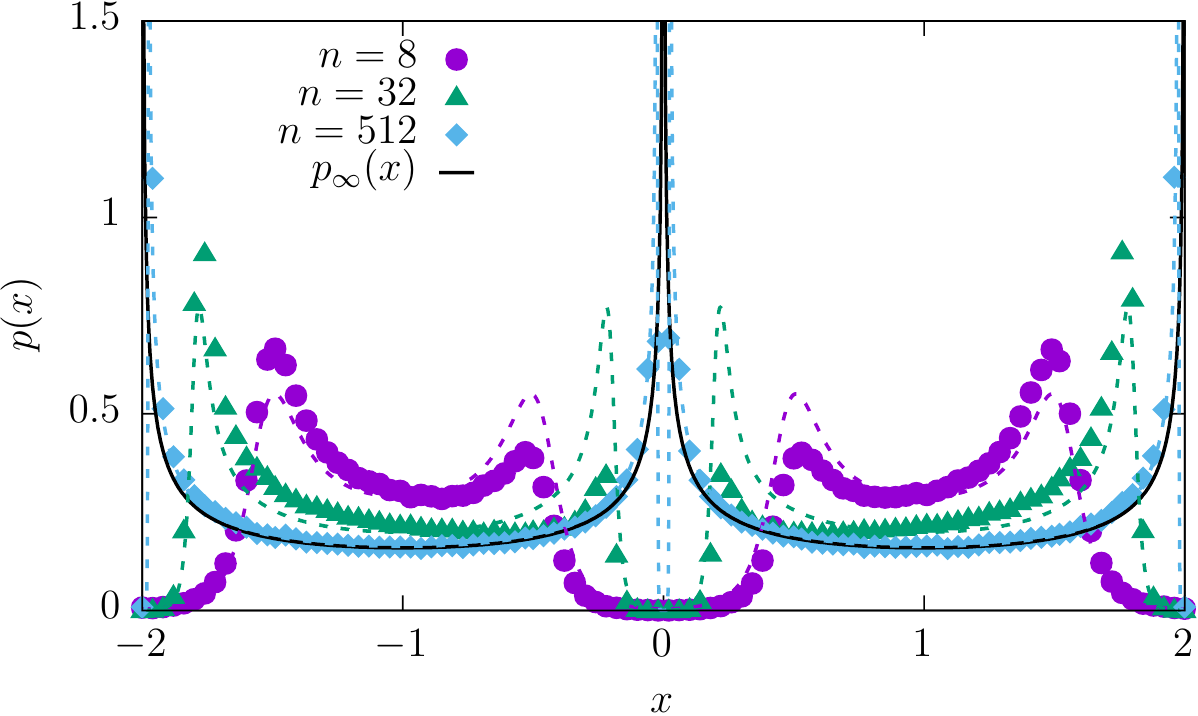} \\
\end{tabular}
\caption{Stationary solutions for $\alpha=1$ with  $n\in\{8,32,512\}$ and $V(x)=n(|x|-1)^n$.
}
\label{fig:twowells8-512}
\end{figure}

Subsequently, we have explored the behaviour of stationary states for the increasing values of the exponent $n$ in Eq.~(\ref{eq:dpotential}).
With the increasing $n$, positions of internal modes shift toward $x=0$.
Finally, in the limit of $n\to\infty$, internal maxima of the stationary density merge, resulting in the amplification of the central mode. 
This phenomenon produces a stationary state with three modal values.
At the same time, external modal values shift towards $x=\pm 1$ and their height increase.
For a large enough steepness (exponent) $n$, the stationary state $p_\infty(x)$ is a sum of stationary states in two rectangular potential wells, see Eq.~(\ref{eq:rec-stationary}).
Analogously like in Fig.~\ref{fig:twowells4}, dashed lines in Fig.~\ref{fig:twowells8-512} present exact solutions given by transformed formulas \cite[Eqs.~(38) and (39)]{dubkov2007}, see Eq.~(\ref{eq:sumofshifted}) for the reference.
A comparison of computer simulations (points) with dashed lines demonstrates how the asymptotic density is reached.
It indicates how the interference of internal modes is responsible for significant deviations from theoretical densities.
Finally, this comparison is especially useful for observing the behaviour of outer modes, since it nicely shows how, with the increasing $n$, outer maxima of probability densities approach impenetrable boundaries arising at $\pm1$.
Here, the qualitative explanation of the observed effect is the same as for the single-well potential.
The only difference is that the random pulses can induce transition over the finite potential barrier separating left and right potential wells.

\subsection{Barrier height\label{sec:hpotential}}

\begin{figure}[!h]%
\centering
\begin{tabular}{c}
\includegraphics[angle=0,width=0.95\columnwidth]{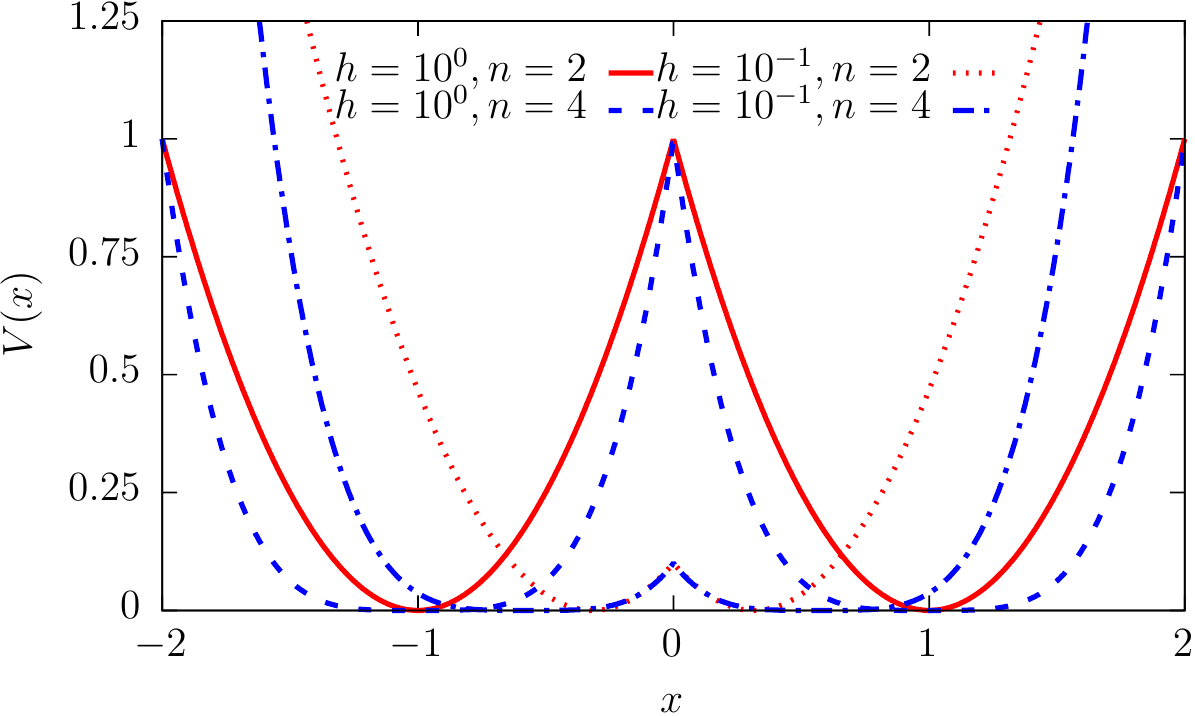} \\
\end{tabular}
\caption{Sample parabolic and quartic potentials, see Eq.~(\ref{eq:h-potential}), with $h=10^0$ and $h=10^{-1}$.
}
\label{fig:h-pot}
\end{figure}

\begin{figure}[!h]%
\centering
\begin{tabular}{c}
\includegraphics[angle=0,width=0.95\columnwidth]{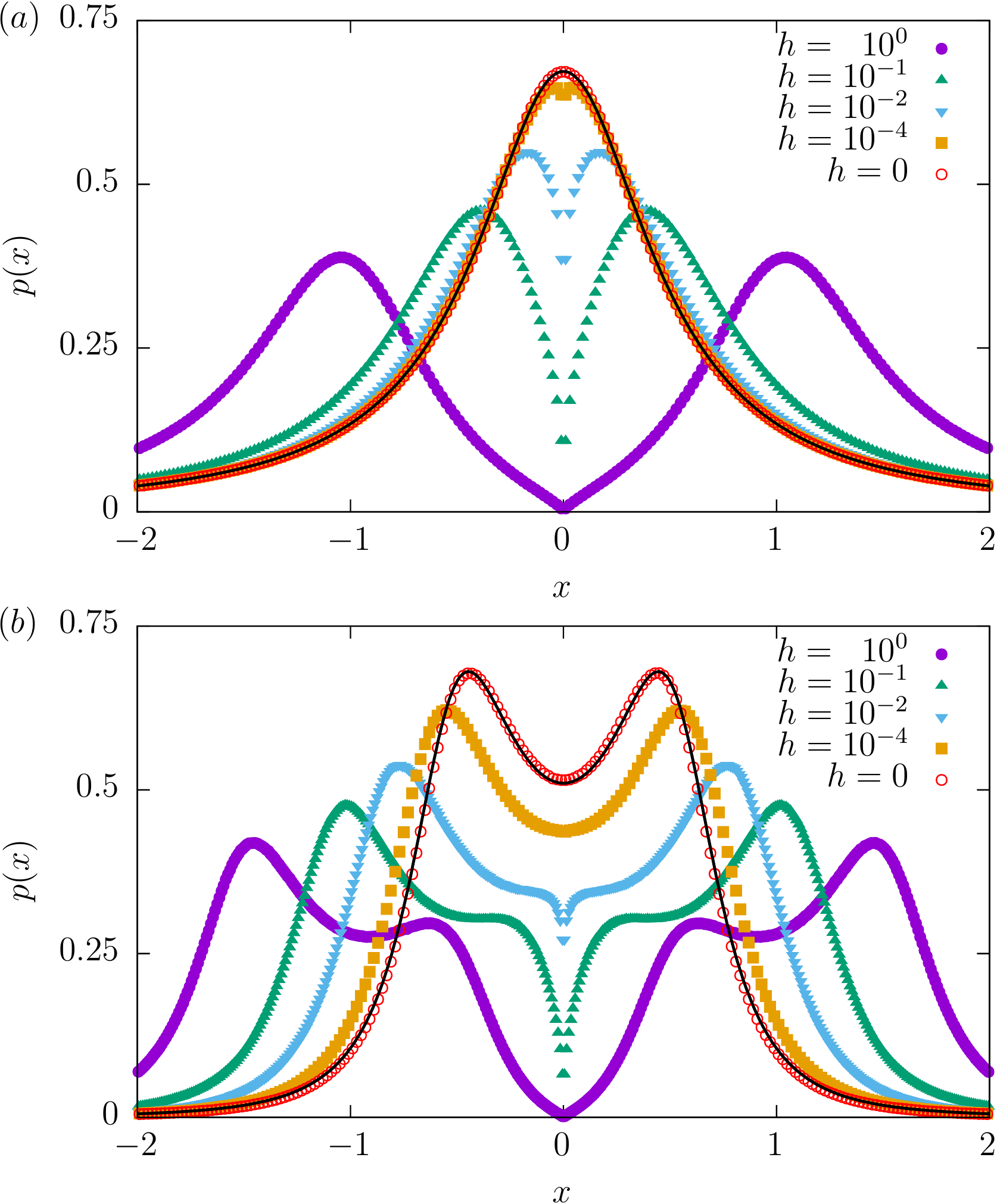} \\
\end{tabular}
\caption{Stationary states for $V(x)=(|x|-h^{1/n})^n$ with $\alpha=1$ and $n=2$ (top panel -- ($a$)) and $n=4$ (bottom panel -- ($b$)).
Black solid lines present $h=0$ theoretical densities, see Eqs.~(\ref{eq:parabolic-h}) and (\ref{eq:quartic-h}).
}
\label{fig:n-h}
\end{figure}

In Sec.~\ref{sec:dpotential}, we have verified how the shape of stationary states changes with the increasing height of the potential barrier separating potential wells.
We have also explored how the limit of double-well infinite rectangular potential well is reached.
Here, we would like to examine the opposite situation.
Namely, we want to investigate how the shape of stationary states changes with the decreasing height of the potential barrier separating both potential wells.
Since the double-well potential given by Eq.~(\ref{eq:dpotential}) is not very suitable for such tests, we use the potential
\begin{equation}
 V(x)=(|x|-h^{1/n})^n,
 \label{eq:h-potential}
\end{equation}
with $n=2$ and $n=4$, see Fig.~\ref{fig:h-pot}.
In Eq.~(\ref{eq:h-potential}), $h$ controls the height of the potential barrier separating minima of the double-well potential. 
The parameter $h$ controls also a position of potential minima.
In the limit of $h=0$, the double-well potential reduces to a single-well potential with a minimum at the origin.
Therefore, for a decreasing $h$, it is possible to see when the barrier separating potential wells is not sufficient to produce maxima of the stationary state in left and right potential wells ($n=2$) or when bimodality within a potential well disappears ($n=4$), see Fig.~\ref{fig:n-h}.

\begin{figure}[!h]%
\centering
\begin{tabular}{c}
\includegraphics[angle=0,width=0.95\columnwidth]{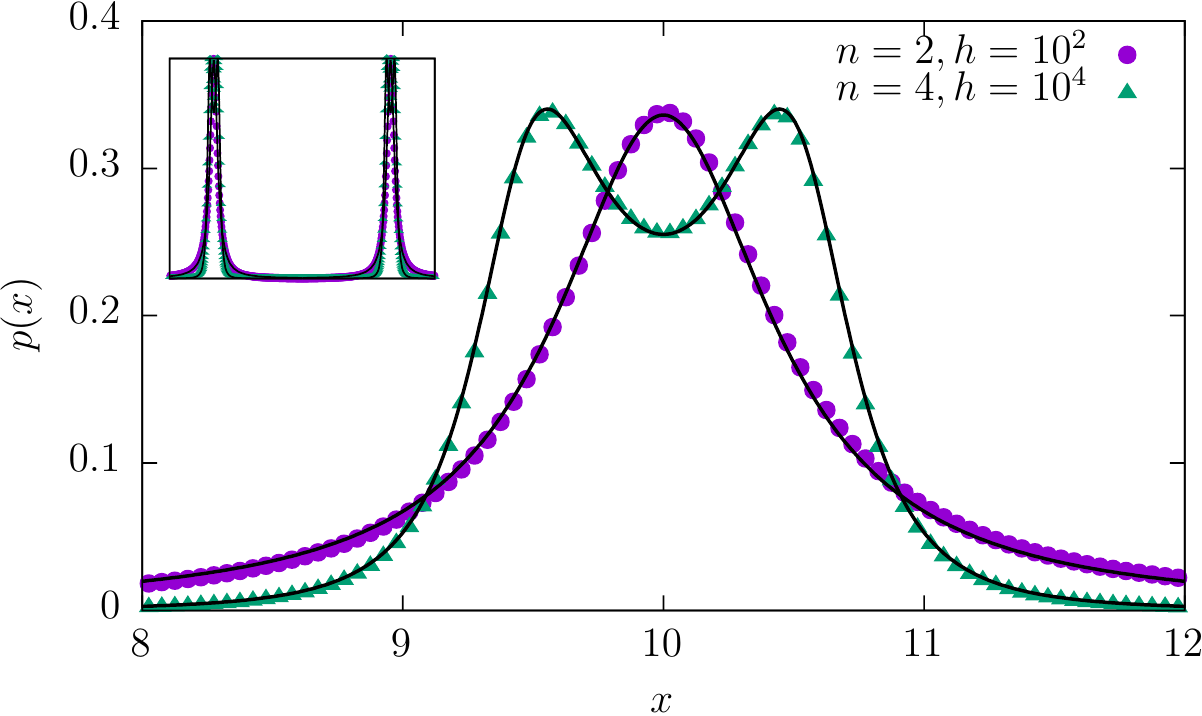} \\
% \missingfigure[figwidth=6cm]{}
\end{tabular}
\caption{Stationary states for $V(x)=(|x|-h^{1/n})^n$ with $\alpha=1$ and $n\in\{2,4\}$. Various curves correspond to various values of $h$.
Black solid lines present shifted and normalized theoretical densities, see Eqs.~(\ref{eq:parabolic-h}) and (\ref{eq:quartic-h}).
The inset shows whole histograms, while the main plot zooms the vicinity of the right potential minimum.
}
\label{fig:n-h-large-h}
\end{figure}

The top panel of Fig.~\ref{fig:n-h} presents results for $V(x)=(|x|-\sqrt{h})^2$.
For $h$ large enough, the potential is built by two parabolas separated by a high potential barrier.
Therefore, within each potential well, there is a stationary state which is given by the stationary state for the harmonic potential, see Fig.~\ref{fig:n-h-large-h}.
With the decreasing $h$, both peaks of the stationary density are approaching each other and start to interfere.
In the limit of $h=0$, the potential is of a single-well, parabolic type.
Consequently, the stationary state is the same as for the stochastic (overdamped) harmonic Cauchy oscillator.
From computer simulations, one may observe that for $h=10^{-4}$ the stationary state is very close to the stationary state for $V(x)=x^2$.
A perfect agreement is obtained for $h=0$. 
However, due to the cusp in the potential, even for a very small $h$, $p(x=0)$ is reduced in comparison to the stationary density for the stochastic harmonic oscillator.
The cusp at $x=0$ moves from the origin some of particles, which want to accumulate there.
Finally, for $h=0$, the single-well parabolic potential  is recovered and the unimodal stationary state is given by the appropriately rescaled Cauchy density, which for $V(x)=x^2$ and $\sigma=1$ is given by Eq.~(\ref{eq:cauchy}) with $\lambda=1/2$, i.e.
\begin{equation}
p(x)=\frac{2}{\pi} \times \frac{1}{(2x)^2+1}.
\label{eq:parabolic-h}
\end{equation}

Similar situation is observed for $n=4$, see Eq.~(\ref{eq:h-potential}) and the bottom panel of Fig.~\ref{fig:n-h}.
For $n=4$ and large enough $h$, the potential is built by two quartic, single-well potentials. 
Therefore, within each potential well, the stationary state is bimodal, see Eq.~(\ref{eq:quartic-stationary}) and Figs.~\ref{fig:spotential} and~\ref{fig:n-h-large-h}.
With the decreasing $h$, internal maxima of the stationary state decrease.
Simultaneously, outer modes shift towards the origin and their height increase.
Finally, for $h=0$, the single-well quartic potential $V(x)=x^4$ is recovered. 
For $\sigma=1$ in the Langevin equation, the bimodal stationary state is given by the equation derived from Eq.~(\ref{eq:transformation-n-s}) with $n=4$ and $\lambda=1/4$, i.e.
\begin{equation}
p(x)= \frac{2}{\pi}  \times \frac{1}{(2x)^4-(2x)^2+1}.
\label{eq:quartic-h}
\end{equation}

The potential given by Eq.~(\ref{eq:h-potential}) is one of many possible potentials that allow to investigate the role of a height of a barrier separating potential's minima on the shape of stationary states. 
The potential (\ref{eq:h-potential}) have been used due to its properties.
In Eq.~(\ref{eq:h-potential}) the $h$ parameter controls the barrier height. 
On the one hand, single well potentials, e.g. harmonic ($n=2$) or quartic ($n=4$), are recovered in the limit of $h\to0$.
On the other hand, for very large $h$, there are no transitions between potential minima and the potential is practically built from two independent single-well potentials.
One of the drawbacks of the potential (\ref{eq:h-potential}) is the fact that the increase in $h$ shifts also positions of potential minima into the direction of larger $|x|$, see Fig.~\ref{fig:h-pot}.
Fig.~\ref{fig:n-h-large-h} presents stationary states for large $h$: $h=10^4$ (quartic potential) and $h=10^2$ (parabolic potential).
For such values of $h$, potential minima are located at $x=\pm 10$, where central parts of within the well probability densities are located.
In every potential well, probability densities perfectly follow shifted and renormalized densities for single well potentials, see Eqs.~(\ref{eq:parabolic-h}) and (\ref{eq:quartic-h}).
The perfect agreement between theoretical single-well densities and numerically estimated histograms originates in the height and width of the potential barrier, which makes the transition from one well to another very unlikely.  
Alternatively, one can generalize potential (\ref{eq:dpotential}) to non-integer $n$ or use $V(x)=h(|x|-1)^n$.
Unfortunately, these generalizations also possess some disadvantages. 
For the generalized Eq.~(\ref{eq:dpotential}), with the change in $n$ not only the barrier height is changed, but also the steepness (exponent) and the ``slope'' (prefactor) of the potential are modified.
In the case of $V(x)=h(|x|-1)^n$, the steepness is fixed but the ``slope'' changes.
Noticeably, in the limit of $h\to0$, this potential does not reduce to the single-well potential, because it disappears.
Nevertheless, we use this potential to inspect the role of multimodality on noise induced effects, see Sec.~\ref{sec:escape}.

Naturally, it is still possible to fine-tailor other, more complicated, potentials bearing required properties.
Nevertheless, we have used potential (\ref{eq:h-potential}) as a simple potential that allows to study the role of the barrier height on the shape of stationary states.
Already, from the potential (\ref{eq:h-potential}) it is possible to draw many general conclusions, see Sec.~\ref{sec:summary}.

%%%%%%%%%%%%%%%%%%%%%%%%%%%%%%%%%%%%%%%%%%%%%%%%%%%%%%%%%%%%%%%%%%%%%%%%%%%%%%%%%%%%%%%%%
%%%%%%%%%%%%%%%%%%%%%%%%%%%%%%%%%%%%%%%%%%%%%%%%%%%%%%%%%%%%%%%%%%%%%%%%%%%%%%%%%%%%%%%%%
%%
%% escape
%%
% \clearpage
\section{Escape from a potential well \label{sec:escape}}

In Sec.~\ref{sec:model} we have inspected conditions for emergence of multiple modes within a single potential well.
In the current section we compare the noise induced escape from the potential well in order to explore the role of a multimodality of stationary states.

\begin{figure}[!h]%
\centering
\begin{tabular}{c}
\includegraphics[angle=0,width=0.95\columnwidth]{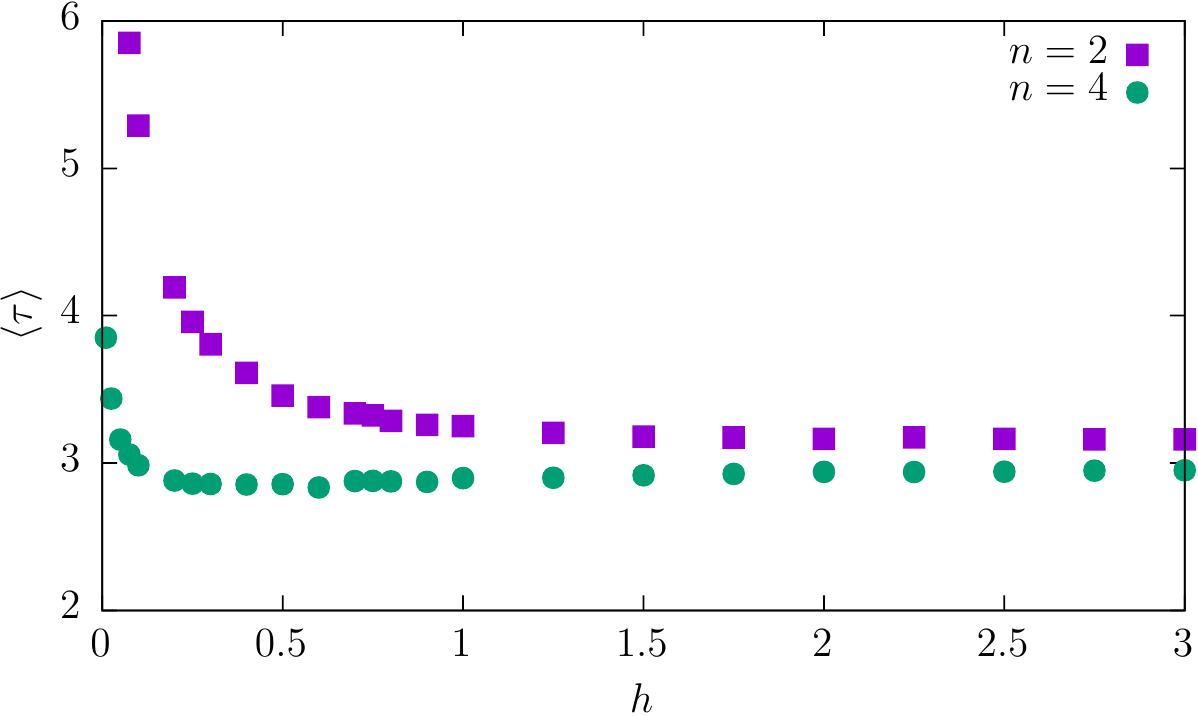}  \\
\end{tabular}
\caption{The mean first passage time for the parabolic and quartic potentials.
Error bars are smaller than the symbol size.
}
\label{fig:mfpt}
\end{figure}

\begin{figure}[!h]%
\centering
\begin{tabular}{c}
\includegraphics[angle=0,width=0.95\columnwidth]{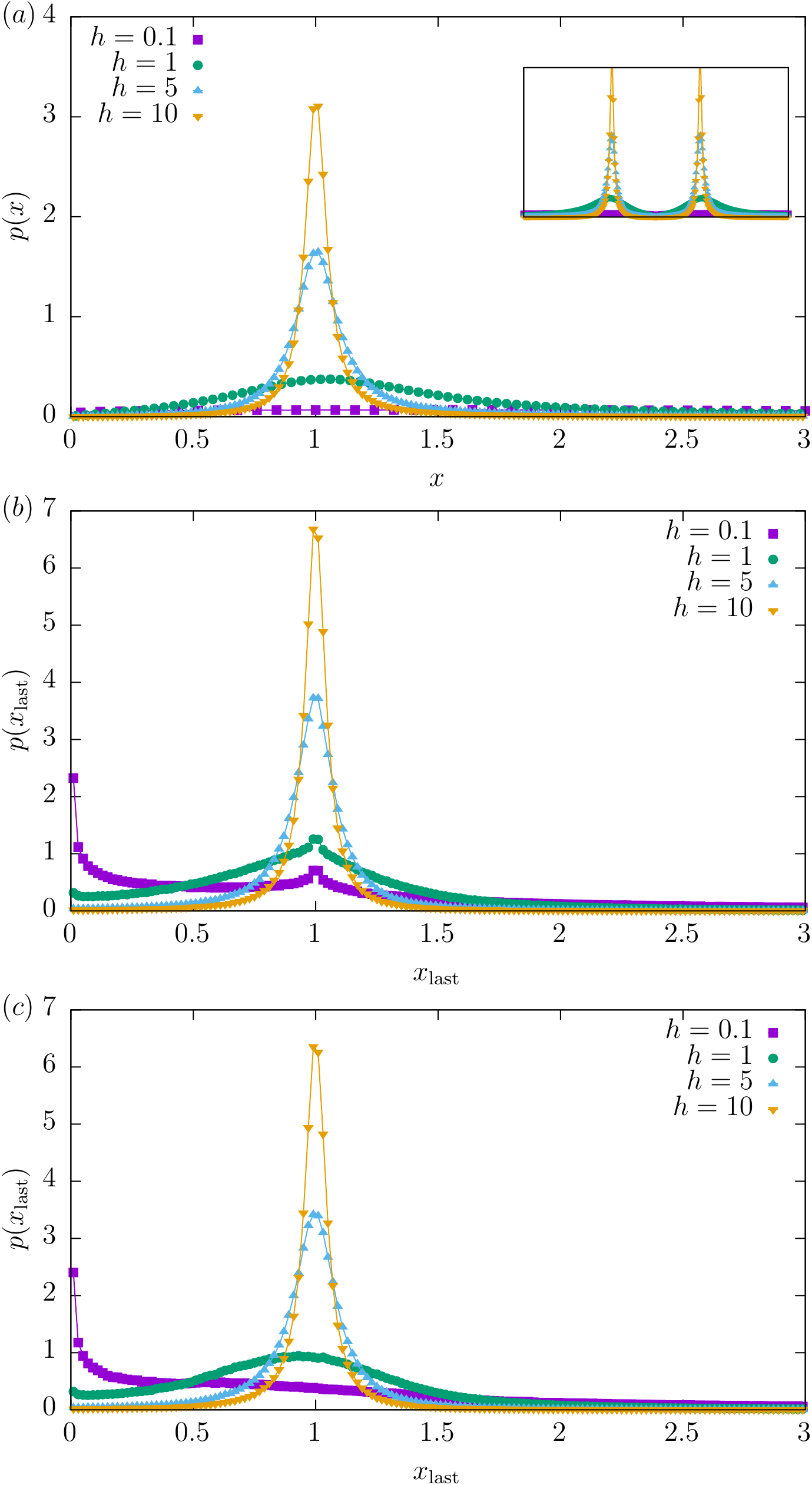} \\
\end{tabular}
\caption{Stationary states (top panel -- ($a$)), last hitting point densities for $x(0)=1$ (middle panel -- ($b$)) and last hitting point densities for $x(0) \sim \mathcal{U}(0.5,1.5)$ (bottom panel -- ($c$)) for $n=2$.
The main plot in the top panel shows the stationary density for $x>0$, while the inset presents the whole density.
}
\label{fig:escape-n2}
\end{figure}

\begin{figure}[!h]%
\centering
\begin{tabular}{c}
\includegraphics[angle=0,width=0.95\columnwidth]{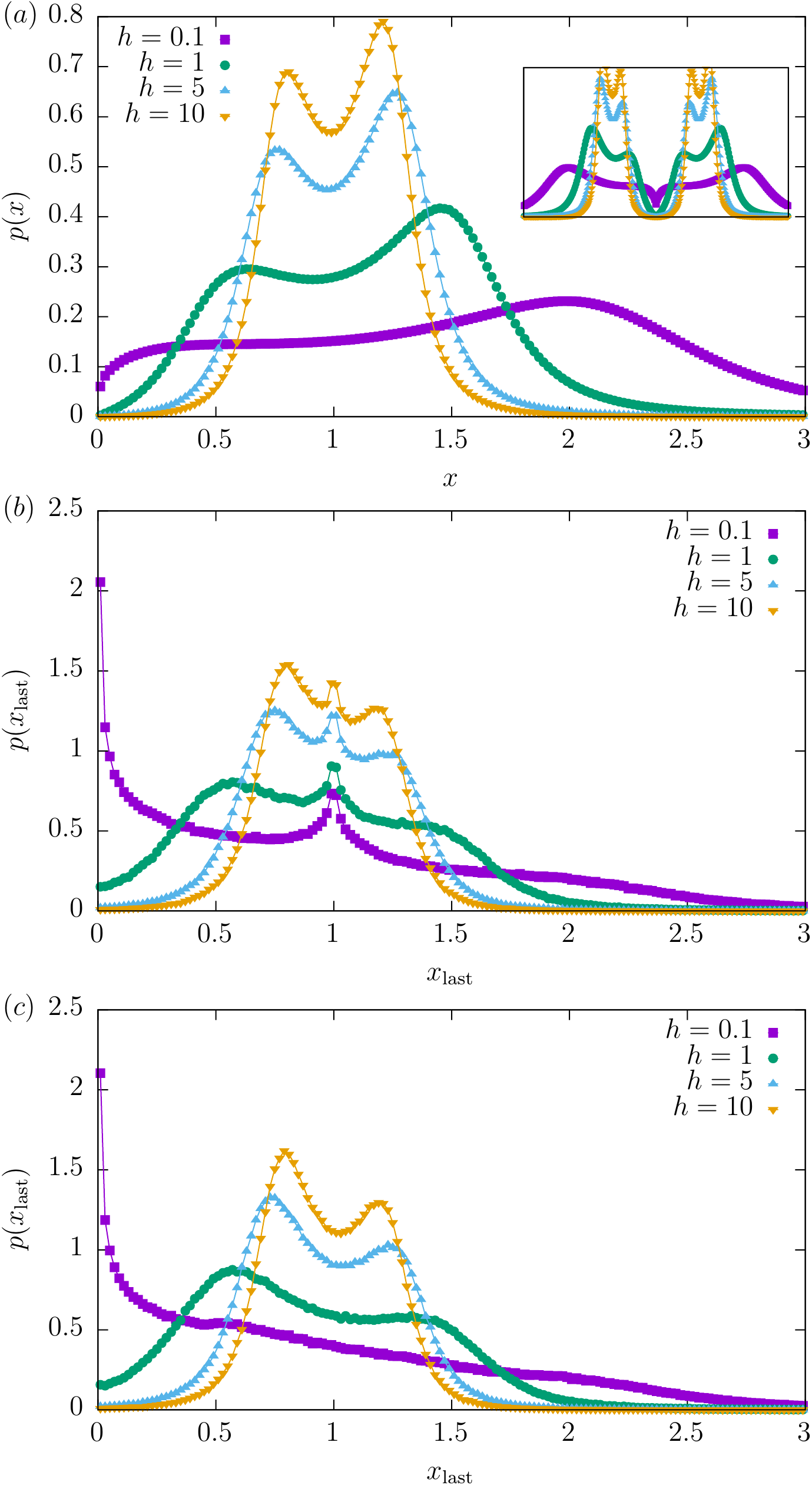} \\
\end{tabular}
\caption{The same as in Fig.~\ref{fig:escape-n2} for $n=4$.
}
\label{fig:escape-n4}
\end{figure}

We consider a motion of the particle driven by the Cauchy noise in the double-well potential 
\begin{equation}
    V(x)=h(|x|-1)^n,
    \label{eq:escape-potential}
\end{equation}
with $n\in\{2,4\}$.
We are studying the escape from a potential well \cite{kramers1940,dybiec2007,chechkin2007,ditlevsen1999}, due to symmetry of the problem, we can assume $x(0)=1$ and consider $x>0$ only.
$x=0$ is the point separating left and right potential wells.
Therefore, the stochastic dynamics is continued as long as $x\geqslant0$, see Appendix~\ref{sec:app-numerical}.
Every first escape from the right potential well is associated with the first passage time
\begin{equation}
    \tau=\min\{\tau : x(0)=1 \;\mbox{and}\; x(\tau) \leqslant 0  \}.
    \label{eq:fpt_def}
\end{equation}
The mean first passage time (MFPT) is the average of first passage times.
In Fig.~\ref{fig:mfpt} the dependence of the MFPT $\langle \tau \rangle$ on the barrier height $h$ is presented.
For the parabolic potential the mean first passage time is larger than for the quartic potential.
With the increasing barrier height ($h>10$) the MFPTs become comparable.
The MFPT is influenced also by size of the domain in which the random walk can be performed, i.e. by the outer ($|x| \gg 1$) parts of the potential.
For the quartic potential, the particle is effectively bounded in a smaller fraction of space than for the parabolic potential, see top panels of Figs.~\ref{fig:escape-n2} and~\ref{fig:escape-n4}.
The effective domain size depends also on the $h$ parameter.
For instance, stationary states for $V(x)=hx^4$ and $V(x)=hx^2$ can be found from  Eq.~(\ref{eq:quartic-n-s}) with $\lambda=h/4$ and from Eq.~(\ref{eq:cauchy})  with $\lambda=h/2$, respectively.
The decrease of the $h$ parameter makes excursions to the large $|x|$ more probable.
In turn these excursion are responsible for the increase in the MFPT.
In the limit of $h\to 0$, a motion in the potential given by Eq.~(\ref{eq:escape-potential}) reduces to the problem of the escape from the positive half-line which is characterized by the diverging mean first passage time \cite{sparre1953,sparre1954,redner2001,chechkin2003b,dybiec2016jpa}.
Therefore, in the limit of $h\to0$ the MFPT $\langle \tau \rangle \to \infty$.
For large $|x|$, the parabolic potential is less steep, consequently, increase in the MFPT with the decrease of the barrier height for the parabolic potential is more rapid than for the quartic potential.
Therefore, the role of a multimodality on escape kinetics can be assessed for a high enough potential barrier.
The top panel of Figs.~\ref{fig:escape-n2} and~\ref{fig:escape-n4} suggest that $h=1$ is sufficient because, in the stationary state, the majority of the probability mass is localized in comparable domains.

Changes in the shape of stationary states, see top panels of Fig.~\ref{fig:escape-n2} and~\ref{fig:escape-n4}, are also reflected in the dependence of the MFPT on the barrier height $h$, see Fig.~\ref{fig:mfpt}.
For $n=2$, the MFPT monotonically decays with the increase of $h$. 
At the same time the maximum of the stationary density does not move and it is located at $x\approx 1$, see top panel of Fig.~\ref{fig:escape-n2}.
For $n=4$, the non-monotonous dependence of the MFPT, with a shallow minimum at $h\approx 0.5$, is due to changes in the shape of the stationary states, see top panel of Fig.~\ref{fig:escape-n4}.

The significant differences between the parabolic and the quartic potential are visible on the level of stationary states, see top panel of Figs.~\ref{fig:escape-n2} and~\ref{fig:escape-n4}.
Analogously, like for the potentials considered in Secs.~\ref{sec:dpotential} and~\ref{sec:hpotential}, if the barrier is high enough, the stationary state in the quartic potential has four modes.
At the same time, the stationary state for the parabolic potential is characterized by two modes located in minima of the potential, i.e. $x=\pm 1$.
In order to elucidate further differences between both potentials we have studied the distribution of the last hitting points.

The last hitting point $x_\mathrm{last}$ is the last point visited before escape from the domain of motion, i.e. it is $x(\tau_i-\Delta t)$, where $\Delta t$ is the integration time step and $\tau_i$ is the first passage time associated with this particular realization of the trajectory.
For the selected setup, it is the last point such that $x>0$.
Due to presence of noise in Eq.~(\ref{eq:langevin}) trajectories are randomized.
Consequently, the first passage time $\tau$ and the last hitting point $x_\mathrm{last}$ are random variables.
The distribution of the last hitting points for both setups is very different, compare middle panels of Figs.~\ref{fig:escape-n2} and~\ref{fig:escape-n4}.
The distribution of last hitting points is related to the stationary state and the initial condition.
A particle is more likely to escape from points where a probability of finding the particle is larger.
Therefore, the last hitting point density follow very different pattern for $n=2$ and $n=4$.
The most probable escape scenario, even for finite $\sigma$, is the escape via a single long jump \cite{ditlevsen1999,imkeller2006,imkeller2006b} because in the weak noise limit the L\'evy noise can be decomposed into the Gaussian white noise and the Poisson compound process.
This does not mean that the particle escapes immediately, because it can be wandering due to the Gaussian component of the noise.
For the parabolic potential, if the barrier has an appropriate height, the escape from the potential well typically is not immediate ($P(\tau > 0.1) > 96\%$) and most probably finishes from the minimum of the potential.
For the quartic potential the situation is more complicated due to two modes of the stationary state within a potential well.
Thus, for the high enough potential barrier, the last hitting point density has three modes: two modes in points where the stationary state has modes and the additional mode in the minimum of the potential, due to immediate escapes from the starting point (initial condition). 
Nevertheless, a immediate escape is not very probable as $P(\tau<0.1)<4\%$, while the integration time step is $10^{-5}$.
The peak arising due to the initial condition can be diminished or removed by the change in the initial condition. For instance for $x(0)$ uniformly sampled from $(0.5,1.5)$, i.e.  $x(0)\sim \mathcal{U}(0.5,1.5)$, there is no third peak, compare middle and bottom panels of Fig.~\ref{fig:escape-n4}.
The height of peaks in the last hitting density is reversed in comparison to the height of peaks in the stationary state, because it is easier to escape from the point which is closer to the boundary.
Moreover, for the decreasing barrier height, the height of peaks located at $|x|\neq 1$ decrease and finally they disappear due to the larger role played by trajectories approaching the boundary separating states.
For a very low potential barrier, regardless of a potential type, the particle can also approach the vicinity of the point separating states, due to ``weak'' noise pulses ruled by the central part of the L\'evy distribution.
Therefore, in addition to the peak in the minimum of the potential, the peak at the boundary emerges.

Lastly, we have compared escape scenarios in both potentials. 
We have estimated probabilities of surmounting the potential barrier via a sequence of given number of jumps into the direction of the boundary.
Tab.~\ref{tab:seq} presents the ratio of probabilities of the passage over the barrier in a single jump $p(1)$ and two consecutive jumps $p(2)$, i.e. $p(1)/p(2)$.
From numerical simulations, it looks that the results included in Tab.~\ref{tab:seq} are robust with respect to the integration time step as long as the integration time step is small enough.
Tab.~\ref{tab:seq} confirms the hypothesis that the main scenario of the L\'evy noise induced escape is an escape via a single long jump \cite{ditlevsen1999,imkeller2006,imkeller2006b,vezzani2018single}.
With the increasing barrier height, the escape via a single long jump becomes dominating.
With the increasing value of the stability index $\alpha$ longer sequences are observed.
Finally, in the limits of $\alpha=2$, i.e. for the Gaussian white noise, the escape takes place in a sequence of jumps.
This effect is fully coherent with the decomposition of the L\'evy noise into the Gaussian white noise and the Poisson compound process \cite{imkeller2006,imkeller2006b,imkeller2010hierarchy}.
In the weak noise limits (small $\sigma$), transitions over the potential barrier are due to Poissonian component of the noise.

\begin{table}[!h]
    \centering
    \begin{tabular}{l||c|c}
          $h$ & $n=2$ & $n=4$ \\ \hline \hline  
                  $0.1$ & 0.75/0.14 & 0.73/0.15 \\ 
          $1$ & 0.82/0.12 & 0.79/0.13 \\ 
          $5$ & 0.86/0.09 & 0.80/0.11\\ 
          $10$ & 0.87/0.09 & 0.81/0.11 \\  
    \end{tabular}
    \caption{The ratio of probabilities of escape in a single  jump and two consecutive jumps, i.e. $p(1)/p(2)$.}
    \label{tab:seq}
\end{table}

%%%%%%%%%%%%%%%%%%%%%%%%%%%%%%%%%%%%%%%%%%%%%%%%%%%%%%%%%%%%%%%%%%%%%%%%%%%%%%%%%%%%%%%%%
%%%%%%%%%%%%%%%%%%%%%%%%%%%%%%%%%%%%%%%%%%%%%%%%%%%%%%%%%%%%%%%%%%%%%%%%%%%%%%%%%%%%%%%%%
%%
%% summary
%%
% \clearpage
\section{Summary and Conclusions\label{sec:summary}}

We have studied stationary states in single-well and double-well potentials.
It has been demonstrated that, for a sufficiently high potential barrier separating minima of double-well potentials, stationary states for systems perturbed by the L\'evy noises can be at least bimodal, due to the possibility of producing more than one maximum of the stationary state within each potential well.
%it is possible to produce more than one maximum of the stationary state.
In order to produce two maxima of the stationary density within a potential well, the potential well needs to be steeper than parabolic. 
For instance, for the double-well potential build from two single-well potentials of $x^4$ type, the stationary density can have four modal values. 
If single-well potentials are of $x^2$ type, the stationary state can be bimodal, with modal values located in potential wells. 
To conclude, the number of modal values is sensitive to the height of the potential barrier separating potential wells and the leading term describing the potential shape in the vicinity of its minima.

In general, it also possible to construct double-well potentials producing more than four modal values of the stationary state. 
Such potentials can be built from two single-well potentials of the type considered in \cite{capala2018multimodal}, i.e. from single-well potentials with non-monotonous dependence of the potential curvature. 
The multimodality of stationary states can be also produced by the  
combined action of the Gaussian white noise and the Markovian  dichotomous noise \cite{dybiec2007c,calisto2017forced}. 
The dynamically induced multimodalidy emerges due to sliding to  minima of the altered potential \cite{dybiec2007c} while the L\'evy noise induced multimodality is produced due to the competition between random long excursion and the deterministic sliding towards minima of the static potential, which is interrupted by long jumps.

Finally, we have studied the role of the multimodality on an escape from a potential well.
We have observed that the escape process from a quartic potential well is slightly faster than the escape from the parabolic potential well, because particles are more likely to concentrate closer to the boundary, due to inner modes of the stationary state.
The concentration of the probability mass closer to the boundary allow for faster escape because a shorter jump is more likely to occur than a longer one.
This effect is the most pronounced for moderate barrier heights.
In addition to stationary states, the most striking differences are visible on the level of last hitting point densities, which follow very different patterns due to different shapes of stationary states.

We have performed our considerations for the Cauchy noise but they can be extended to other L\'evy noises. 
On the one hand, such an extension is straight forward. 
On the other hand, we are not expecting to observe different results (except weakening of effects for $1<\alpha<2$) that one described within the current manuscript.
We have used the Cauchy density due to analytically known stationary states, which can be compared with numerical simulations.

%%%%%%%%%%%%%%%%%%%%%%%%%%%%%%%%%%%%%%%%%%%%%%%%%%%%%%%%%%%%%%%%%%%%%%%%%%%%%%%%%%%%%%%%%
%%%%%%%%%%%%%%%%%%%%%%%%%%%%%%%%%%%%%%%%%%%%%%%%%%%%%%%%%%%%%%%%%%%%%%%%%%%%%%%%%%%%%%%%%
%%
%% appendix 1
%%
% \clearpage
\appendix
\section{Stationary state for $V(x)={x^n}/{n}$ with $n=4$ \label{sec:app-n4} and $\sigma \neq 1$}

The quartic oscillator $V(x)=\frac{x^4}{4}$ is described by the following Langevin equation is
\begin{equation}
 \frac{dx}{dt} = -x^3 + \zeta_\alpha(t).
 \label{eq:c1}
\end{equation}
As proved in \cite{chechkin2002,chechkin2003,chechkin2004,chechkin2006,chechkin2008introduction}, for $\alpha=1$, the stationary state is 
\begin{equation}
 f_{\alpha=1}(x)=  \frac{1}{\pi (x^4-x^2+1)}.
 \label{eq:c1-stationary}
\end{equation}
The stationary state for 
\begin{equation}
 \frac{dx}{dt} = -x^3 + \sigma \zeta_1(t)
 \label{eq:c-sigma}
\end{equation}
can be re-constructed by rescaling the space variable $x$ and time $t$.
Let us introduce
\begin{equation}
 \tilde{x}=\frac{x}{x_0}
\end{equation}
and
\begin{equation}
 \tilde{t}=\frac{t}{t_0}.
\end{equation}
The non-trivial part is the noise transformation
\begin{eqnarray}
\sigma \zeta_\alpha (t) & =  & \sigma \frac{dL(t)}{dt} 
= \sigma \frac{d}{dt}L(t_0\tilde t)
= \sigma \frac{d}{dt}t_0^{\frac{1}{\alpha}}L(\tilde t) \\ \nonumber
&=& \sigma t_0^{\frac{1}{\alpha}} \frac{d}{dt} L(\tilde t)
 = \sigma t_0^{\frac{1}{\alpha}} \frac{d\tilde{t}}{dt}  \frac{d}{d\tilde{t}} L(\tilde t) \\ \nonumber
 & = & \sigma t_0^{\frac{1}{\alpha}-1} \zeta_\alpha(\tilde{t}), 
\end{eqnarray}
where $L(t)$ is the $\alpha$-stable L\'evy motion, which is $\frac{1}{\alpha}$  self-similar process, whose derivative is the $\alpha$-stable (L\'evy) noise $\zeta_\alpha(t)$.
Eq.~(\ref{eq:c-sigma}) transforms into
\begin{equation}
 \frac{x_0}{t_0} \frac{d\tilde{x}}{d\tilde{t}} = -x_0^3 \tilde{x}^3 + \sigma t_0^{\frac{1}{\alpha}-1} \zeta_\alpha(\tilde{t}).
 \label{eq:transformed1}
\end{equation}
First, we multiply both sides of Eq.~(\ref{eq:transformed1}) by $\frac{t_0}{x_0}$.
Next, we require that prefactors in the deterministic force and in the random force (noise) are equal to unity
\begin{equation}
 \left\{
 \begin{array}{lcl}
 t_0 x_0^2 & = & 1 \\
 \sigma \frac{ t_0^{\frac{1}{\alpha}} }{x_0} & = & 1 \\
 \end{array}.
 \right.
\end{equation}
For $\alpha=1$, the solution of the above equation is
\begin{equation}
 \left\{
 \begin{array}{lcl}
 x_0  & = & \sigma^{\frac{1}{3}} \\
 t_0  & = & \sigma^{-\frac{2}{3}} \\
 \end{array}.
 \right.
\end{equation}
After rescaling of variables Eq.~(\ref{eq:c-sigma}) takes the form of Eq.~(\ref{eq:c1}) with $x$ and $t$ interchanged with $\tilde{x}$ and $\tilde{t}$.
Consequently, the stationary state is given by Eq.~(\ref{eq:c1-stationary}) with $x$ replaced by $\tilde{x}$ because the rescaling of time does not change the stationary state.
The stationary state for $x$ can be reconstructed by the change of variables 
\begin{eqnarray}
 f_{\alpha=1}(x) & = &  f_{\alpha=1}\left(\tilde{x} = \frac{x}{x_0} \right) \times \frac{1}{x_0} \\ \nonumber
 & = &  f_{\alpha=1}\left(\tilde{x} = \frac{x}{\sigma^{\frac{1}{3}}} \right) \times \frac{1}{\sigma^{\frac{1}{3}}} \nonumber \\
 & = & \frac{1}{\pi \sigma^{\frac{1}{3}} } \times \frac{1}{\left[\frac{x}{\sigma^{\frac{1}{3}}}\right]^4-\left[\frac{x}{\sigma^{\frac{1}{3}}}\right]^2+1}. \nonumber
\end{eqnarray}

%%%%%%%%%%%%%%%%%%%%%%%%%%%%%%%%%%%%%%%%%%%%%%%%%%%%%%%%%%%%%%%%%%%%%%%%%%%%%%%%%%%%%%%%%
%%%%%%%%%%%%%%%%%%%%%%%%%%%%%%%%%%%%%%%%%%%%%%%%%%%%%%%%%%%%%%%%%%%%%%%%%%%%%%%%%%%%%%%%%
%%
%% appendix 2
%%
\section{Stationary states for $V(x)=\lambda n x^n$ with $n=4$ and $n=2$ \label{sec:app-n}}

For $V(x)=\lambda n x^n$, the Langevin equation takes the following form
\begin{eqnarray}
\label{eq:nforce}
 \frac{dx}{dt} & = & -V'(x)+\sigma \zeta_\alpha(t) \\
 & = & -\lambda n^2 x^{n-1} +\sigma \zeta_\alpha(t). \nonumber
\end{eqnarray}
Please note that the scale parameter of noise $\zeta_\alpha$ is set to 1.
Thus $\sigma \zeta_\alpha$ acts as the $\alpha$-stable noise with the scale parameter $\sigma$.
By the appropriate change of variables, the prefactor in the deterministic force can be incorporated into the $\sigma$ parameter allowing us to use results of Appendix~\ref{sec:app-n4}.
Let us substitute $x=ay$ in Eq.~(\ref{eq:nforce})
\begin{equation}
 a \frac{dy}{dt} = -\lambda a^{n-1} n^2 y^{n-1} + \sigma \zeta_\alpha
\end{equation}
and divide it by $a$
\begin{equation}
  \frac{dy}{dt} = -\lambda a^{n-2} n^2 y^{n-1} + \frac{\sigma}{a} \zeta_\alpha.
  \label{eq:divided}
\end{equation}
For $n\neq 2$, the value of $a$ is determined by the condition
\begin{equation}
\lambda a^{n-2}n^2 = 1 
\end{equation}
resulting in
\begin{equation}
a = (\lambda n^2)^{\frac{1}{2-n}}.
\end{equation}
Therefore, Eq.~(\ref{eq:divided}) transforms into
\begin{equation}
  \frac{dy}{dt} = - y^{n-1} + \sigma' \zeta_\alpha,
  \label{eq:transformed}
\end{equation}
where 
\begin{equation}
 \sigma'= \frac{\sigma}{a} = \frac{\sigma}{(\lambda n^2)^{\frac{1}{2-n}}}.
\end{equation}
Consequently, from the stationary  solution $f(y)$ of Eq.~(\ref{eq:transformed}) the stationary state $f(x)$ of Eq.~(\ref{eq:nforce}) can be obtained by the transformation of variables
\begin{equation}
 f(x) = f\left(y=\frac{x}{a} \right) \times \frac{1}{a}. 
 \label{eq:transformation-n-s}
\end{equation}

In particular for $n=4$, $\lambda=1$ and $\alpha=1$ we have $a=\frac{1}{4}$ and $\sigma'=4\sigma$.
In the special case of the quartic Cauchy oscillator the stationary solution is
\begin{eqnarray}
\label{eq:quartic-n-s}
 f(x) & = &  \frac{4}{\pi (\sigma')^{\nicefrac{1}{3}}} \times \frac{1}{\left[ \frac{y}{(\sigma')^{\nicefrac{1}{3}}}\right]^4-\left[\frac{y}{(\sigma')^{\nicefrac{1}{3}}}\right]^2+1} \Bigg|_{\sigma'=4\sigma,y=4x} \\
  & = &  \frac{4}{\pi (4\sigma)^{\nicefrac{1}{3}}} \times \frac{1}{\left[ \frac{4x}{(4\sigma)^{\nicefrac{1}{3}}}\right]^4-\left[\frac{4x}{(4\sigma)^{\nicefrac{1}{3}}}\right]^2+1}. \nonumber  
\end{eqnarray}

Similar calculations can be performed for any $n>0$, however for $n=2$ one needs to transform both space and time in the full Langevin equation. Following calculations from Appendix~\ref{sec:app-n4}, for $n=2$ with $\alpha=1$, one obtains $t_0=1/4\lambda$ and $x_0=\sigma/4\lambda$.
Consequently, the stationary state is the Cauchy density
\begin{eqnarray}
\label{eq:cauchy}
%  f(x)=f\left(y=\frac{x}{x_0} \right) \times \frac{1}{x_0},
 f(x) & = & \frac{1}{\pi} \frac{1}{y^2+1} \times \frac{1}{x_0} \Bigg|_{y=x/x_0} \\
 & = &  \frac{1}{\pi} \frac{1}{(4\lambda x/\sigma)^2+1} \times \frac{4\lambda }{\sigma}. \nonumber
\end{eqnarray}
% which for $\sigma=1$ and $\lambda=1/2$ reduces to 
% \begin{equation}
%  f(x)=\frac{2}{\pi} \times \frac{1}{(2x)^2+1}.
% \end{equation}

%%%%%%%%%%%%%%%%%%%%%%%%%%%%%%%%%%%%%%%%%%%%%%%%%%%%%%%%%%%%%%%%%%%%%%%%%%%%%%%%%%%%%%%%
%%%%%%%%%%%%%%%%%%%%%%%%%%%%%%%%%%%%%%%%%%%%%%%%%%%%%%%%%%%%%%%%%%%%%%%%%%%%%%%%%%%%%%%%
%
% appendix 3
%
\section{Numerical methods\label{sec:app-numerical}}

The Langevin equation, see Eq.~(\ref{eq:langevin})
\begin{equation}
    \frac{dx}{dt}=-V'(x)+\sigma \zeta_\alpha(t)
\end{equation}
is integrated with the Euler-Maryuama scheme
\begin{eqnarray}
\label{eq:em}
 \Delta x & = & x(t+\Delta t)-x(t) \\
 & = & -V'(x(t))\Delta t +\Delta t^{1/\alpha} \sigma \zeta_t \nonumber,
\end{eqnarray}
where $\zeta_t$ represents independent, identically distributed random variables  \cite{chambers1976,weron1995,weron1996} following the symmetric $\alpha$-stable density \cite{janicki1994,janicki1996}.
The Euler-Maryuama scheme assures proper interpretation of stochastic integrals \cite{burrage2006comment}, which become more complex due to discontinuity of L\'evy flights.
Eq.~(\ref{eq:em}) is accompanied with the initial condition on $x(0)$ which can be deterministic, i.e. $x(0)=x_0$, or random, e.g. $x(0) \sim \mathcal{U}(a,b)$.
In the absence of impenetrable boundaries, stationary states are independent to the selection of $x(0)$, while for the first escape problems results are usually sensitive to the choice of $x(0)$.

Scheme~(\ref{eq:em}) is used both for the construction of stationary states and examination of first escape problems.
Crucial parameters in the approximation~(\ref{eq:em}) are the integration time step $\Delta t$ and the number of repetitions $N$. The number of repetitions controls the statistical error of the approximation and fluctuations of the histogram.
The integration time step controls the systematic error.
These parameters are either selected by comparison of results of computer simulations with theoretical predictions or by self-consistency tests.
Therefore, we have performed a series of simulations with the decreasing integration time step.
This sequence was stopped when results were coherent with theoretical predictions or statistically equivalent to results with a larger integration time step.
After fixing $\Delta t$ we have increased the number of repetitions in order to decrease statistical errors.
In order to asses whether a stationary state was reached we have measured the interquantile widths. Simulations were performed till the time $t_\mathrm{max}$ which is long enough to assure that the interquantile width is not increasing.
The interquantile width is more robust than the standard deviation because it can be defined also for distributions with the diverging variance.

The problem of first escape is also studied with the help of Eq.~(\ref{eq:em}).
After fixing $x(0)$ the simulation is performed as long as the particle has not crossed the prescribed boundary. Here, as long as $x>0$ .
From many simulations of escape events we have recorded a series of first passage times $\tau_i$ and last hitting points $x_\mathrm{last}$, i.e. positions of the last visited point before leaving the domain of motion. In the next step from these series we can calculate the mean first passage time $\langle \tau \rangle$  along with its error and estimate statistics of last hitting points, e.g. $p(x_{\mathrm{last}})$.

Within simulations the integration time step varied between $10^{-3}$ -- $10^{-5}$, while the number of repetitions was adjusted to  $10^5$  -- $10^8$.
Stationary states were constructed with a smaller integration time step and a larger number of repetitions because they were constructed using CUDA on GPU (graphics cards).
The MFPT was estimated on conventional CPUs, consequently statistics was poorer than for stationary states.

%%%%%%%%%%%%%%%%%%%%%%%%%%%%%%%%%%%%%%%%%%%%%%%%%%%%%%%%%%%%%%%%%%%%%%%%%%%%%%%%%%%%%%%%%
%%
%% ACKNOWLEDGMENTS
%%

\begin{acknowledgments}
This project was supported by the National Science Center grant (2014/13/B/ST2/02014).
This research was supported in part by PLGrid Infrastructure.
Computer simulations have been performed at the Academic
Computer Center Cyfronet, AGH University of Science and Technology (Krak\'ow, Poland)
under CPU grant DynStoch.

\end{acknowledgments}

% \bibliography{core-bibliography,bibliography}

\def\url#1{}

\end{document}